\documentclass[10pt,journal,final,twocolumn]{IEEEtran}
\usepackage{amsmath,amsfonts}
\usepackage{algorithmic}
\usepackage{algorithm}
\usepackage{array}
\usepackage[caption=false,font=normalsize,labelfont=sf,textfont=sf]{subfig}
\usepackage{textcomp}
\usepackage{stfloats}
\usepackage{url}
\usepackage{verbatim}
\usepackage{graphicx}
\usepackage{cite}
\usepackage{color}
\usepackage{mathrsfs}
\usepackage{bm}
\usepackage{booktabs}
\usepackage{pifont}     
\usepackage{chngcntr}   

\usepackage{ulem}
\usepackage{balance}

\usepackage{makecell} 
\usepackage{threeparttable} 

\begin{document}

\title{Energy-efficient Time-modulated Beam-forming for Joint Communication-Radar Systems}

\author{Chengzhao~Shan,~\IEEEmembership{Member,~IEEE,}~Jiayan~Zhang,~Yongkui~Ma,~\IEEEmembership{Member,~IEEE,}~Xuejun~Sha,~\IEEEmembership{Member,~IEEE,}
Honglin~Zhao,~\IEEEmembership{Member,~IEEE},~Jiankang~Zhang,~\IEEEmembership{Senior~Member,~IEEE,}~Lajos~Hanzo,~\IEEEmembership{Life~Fellow,~IEEE}
\thanks{Manuscript received XXX; revised XXX.
This work was supported by the National Natural Science Foundation of China under Grants 62171151. Corresponding author: H. Zhao and L. Hanzo.}
\thanks{C. Shan, X. Sha and H. Zhao are with School of Electronics and Information Engineering, Harbin Institute of Technology (HIT), Harbin 150001, China, and also with the Science and Technology on Communication Networks Laboratory, Shijiazhuang 050080, China (E-mails:\{czshan, shaxuejun, hlzhao\}@hit.edu.cn).}
\thanks{J. Zhang and Y. Ma are with School of Electronics and Information Engineering, Harbin Institute of Technology (HIT), Harbin 150001, China (E-mails:\{jyzhang, yk\_ma\}@hit.edu.cn).}
\thanks{J. Zhang is with Department of Computing \& Informations, Bournemouth University, BH12 5BB, U.K. (E-mail: jzhang3@bournemouth.ac.uk).}
\thanks{L. Hanzo is with School of Electronics and Computer Science, University of Southampton, SO17 1BJ, U.K. (E-mail: lh@ecs.soton.ac.uk).}}

\markboth{IEEE TRANSACTIONS ON GREEN COMMUNICATIONS AND NETWORKING,~Vol.~XX, No.~XX, AUGUST~2023}%
{Shell \MakeLowercase{\textit{et al.}}: A Sample Article Using IEEEtran.cls for IEEE Journals}


\maketitle

\begin{abstract}
To alleviate the shortage of spectral resources as well as to reduce the weight, volume, and power consumption of wireless systems, joint communication-radar (JCR) systems have become a focus of interest in both civil and military fields. JCR systems based on time-modulated arrays (TMAs) constitute an attractive solution as a benefit of their high degree of beam steering freedom, low cost, and high accuracy. However, their sideband radiation results in energy loss, which is an inherent drawback. Hence the energy-efficiency optimization of TMA-based JCR systems is of salient importance, but most of the existing TMA energy-efficiency optimization methods do not apply to JCR systems. To circumvent their problems, a single-sideband structure is designed for flexibly reconfigurable energy-efficient TMA beam steering. First, some preliminaries on single-sideband TMAs are introduced. Then, a closed-form expression is derived for characterizing the energy efficiency. Finally, the theoretical results are validated by simulations.
\end{abstract}


\section{Introduction}
\IEEEPARstart{W}{ith} the development of cost-efficient consumer electronic technology, the number of wireless communication devices in operation has exploded, which resulted in an impending wireless spectrum crunch \cite{JCR2,TGCN1}. On the other hand, radar and communication systems benefit from continued miniaturization and operation in high-frequency bands. Hence numerous compelling application scenarios have emerged. For civilian use, applications such as the Internet of Vehicles \cite{TVT2}, autonomous driving \cite{TVT3}, and smart homes are widely studied based on the joint design of communication and sensing. From a military perspective, the integration of radar and communication technologies is expected to revolutionize combat systems and to improve the system performance in terms of energy consumption, reduced form-factor, concealment, compatibility, and other aspects \cite{JCR3}. Joint communication-radar (JCR) technology, which combines these two technologies into a single platform, facilitates both hardware reuse and signal waveform sharing. This leads to various advantages, including a low cost, a compact size, and high spectral efficiency \cite{JCR4}.

Generally, JCR designs can be divided into two categories: 1) radar-communication coexistence (RCC) designs and 2) dual-functional radar-communication (DFRC) designs \cite{JCR5}. RCC designs focus on the joint operation of independent radar and communication systems to achieve integration and avoid mutual interference between the two subsystems, as in \cite{JCR6} and \cite{JCR7}. In DFRC designs \cite{TGCN2}, both radar detection and data transmission are performed through a common integrated platform, which naturally achieves full cooperation. Furthermore, DFRC designs can be either communication-centric, radar-centric, or joint designs \cite{JCR2}. Communication-centric designs rely on existing communication signals, such as IEEE 802.11ad \cite{JCR8} and IEEE 802.11p \cite{JCR9}, to achieve radar detection functions, while radar-centric designs embed communication information into existing radar waveforms, such as frequency-modulated continuous waves (FMCW) \cite{JCR10} and linear frequency-modulated (LFM) waves \cite{JCR11}. Joint designs offer a tunable trade-off between communication and radar performance that is not limited by any existing standards; existing examples include designs based on mutual information (MI) optimization \cite{JCR12} and JCR beamforming (BF)\cite{JCR13}.

From the perspective of the antenna configuration on the radio frequency (RF) side, JCR designs include single-antenna and multi-antenna designs. A single-antenna configuration is easy to implement due to its low hardware complexity \cite{JCR14}, but its spatial freedom is limited; thus, the radar and communication functionalities cannot work in different directions simultaneously in this configuration. An important feature of a JCR system is the integration of radar and communication in the airspace \cite{TGCN3}, which benefits from the high degree of beam steering freedom offered by a multi-antenna configuration, such as in \cite{JCR15} and \cite{JCR16}. Many studies have investigated the beam steering of array antennas to realize the integration of radar and communication technology; an important branch of the related research focuses on JCR systems based on time-modulated arrays (TMAs) \cite{JCR17}.

The time dimension was first introduced into array antennas in \cite{TMA1}, and the term TMA first appeared in \cite{TMA2}. With the development of high-speed RF switches, TMAs have received considerable attention and have found various applications, such as adaptive BF \cite{TMA3}, multiple-input multiple-output (MIMO) radar \cite{TMA4}, and direction-finding \cite{TMA5}. A TMA has many benefits \cite{TMA_a1}: 1) its beam steering freedom increases due to the added time dimension, 2) the use of RF switches instead of traditional phase shifters leads to a reduction in cost, and 3) time as a control variable is more accurate than traditional phase shifters. Recently, many TMA-based JCR systems have emerged. Euziere \emph{et al.} proposed a DFRC TMA \cite{TMA6,TMA7,TMA8}, in which the stable main beam is used for radar detection and sidelobe variations are used for communication through amplitude modulation (AM) or 4-state quadrature amplitude modulation (4-QAM). Ahmed \emph{et al.} \cite{TMA9} proposed a DFRC scheme to embed QAM-modulated information into radar waveforms by employing sidelobe control and waveform diversity. However, although this scheme supports the simultaneous operation of radar and communication in different directions, it is difficult to configure the radar beam to scan freely, and the communication beam is fixed. In addition, the communication modulation method is relatively simple and has low reliability. To address the above shortcomings, in \cite{TMA10}, we proposed an integrated radar-communication system based on different TMA harmonic components, where the fundamental component and the 1st upper harmonic component are used for radar scanning and wireless communication, respectively.

Although TMAs are widely applied, they also have the inherent disadvantage that the generation of unwanted harmonic components leads to out-of-band energy leakage. Yang \emph{et al.} \cite{TMA11} studied the problem of energy-efficiency optimization for TMAs  and inspired much follow-on research. Since this optimization problem is nonconvex, most of the existing literature is based on evolutionary algorithms, such as differential evolution (DE) \cite{TMA12}, simulated annealing (SA) \cite{TMA13}, genetic algorithms (GAs) \cite{TMA14}, and various other algorithms \cite{TMA18,EE10,EE0}. There are also some optimization algorithms based on closed-form expressions for the TMA energy losses \cite{Close1,Close2,Ref2}, as detailed in \cite{TMA19} and \cite{Ref1}. In addition, many new structures and modulation modes have been conceived for sideband radiation reduction or energy efficiency optimization \cite{EE13,EE14,EE15,EE22}. However, all the algorithms mentioned above (\cite{TMA11,TMA12,TMA13,TMA14,TMA18,EE10,EE0,TMA19,Ref1,EE13,EE14,EE15,EE22}) retain only the fundamental component and suppress all harmonic components. The goal is to make full use of the ultralow sidelobes of a TMA's fundamental beam pattern, which however conflicts with the demand for multi-harmonic components in the JCR scenario \cite{DE1}. To circumvent the above problems, some of the authors proposed an energy-efficiency optimization algorithm for TMA-based JCR systems \cite{DE1} and presented a quantitative analysis of such a JCR system in \cite{JSTSP1} to facilitate further energy efficiency improvements.

\begin{table*}
\label{TabA1}
\renewcommand{\arraystretch}{1.3}
\centering
\caption{Comparison the main contribution of TMA beam steering method}
\begin{threeparttable}
\begin{tabular}{|l|c|c|c|c|c|c|c|c|}
\hline
  & \cite{TMA6,TMA7,TMA8,TMA9} & \cite{TMA10} & \cite{TMA11,TMA12,TMA13,TMA14,TMA18,EE10,EE0} &\cite{TMA19,EE13,EE14,EE15,EE22}&\cite{Ref1} &\cite{DE1} &\cite{JSTSP1} & Proposed \\
\hline
Applied in JCR & \checkmark & \checkmark &  &  &  & \checkmark & \checkmark & \checkmark \\
\hline
Flexible beam pointing &  & \checkmark & \checkmark &  \checkmark & \checkmark & \checkmark & \checkmark & \checkmark \\
\hline
\makecell[l]{Arbitrary radar/communication\vspace{-3pt}\\ waveform} &  & \checkmark & $\triangle$ & $\triangle$ & $\triangle$ & \checkmark & \checkmark & \checkmark \\
\hline
Energy efficiency optimized &  &  & \checkmark & \checkmark & \checkmark & \checkmark & \checkmark & \checkmark \\
\hline
\makecell[l]{Energy efficiency optimized\vspace{-3pt} \\ (quantitative analysis)} &  &  &  & \checkmark & \checkmark &  & \checkmark & \checkmark \\
\hline
Single-beam system &  &  & \checkmark & \checkmark & \checkmark &  &  & \checkmark \\
\hline
Dual-beam system & \checkmark & \checkmark &  &  &  & \checkmark & \checkmark & \checkmark \\
\hline
Linear array & \checkmark & \checkmark & \checkmark & \checkmark & & \checkmark & \checkmark & \checkmark\\
\hline
Planar array & & & & & \checkmark & & & \checkmark\\
\hline
Volumetric (conformal) array & & & & & & & & \checkmark\\
\hline
\makecell[l]{Reconfigurable single/dual-beam\vspace{-3pt}\\ system} &  &  &  &  &  &  &  & \checkmark \\
\hline
\end{tabular}
    \begin{tablenotes}
        \item $\triangle$: Not involved  
    \end{tablenotes}
\end{threeparttable}
\end{table*}

To sum up, at the time of writing the following issues persist in the TMA-based JCR systems. 1) The beam direction is fixed, the communication mode is a single one, and the harmonic leakage is not considered. 2) Existing TMA energy efficiency optimization methods cannot be directly applied to multi-beam JCR systems. 3) Lack of energy-efficient TMA single/dual-beam reconfigurable configuration scheme. Inspired by the aforementioned observations, this paper further extends the previous studies in \cite{TMA10,DE1,JSTSP1} where a single-sideband structure is utilized for the TMA for completely removing most of the harmonic components. Explicitly, a closed-form expression is derived for the energy efficiency of a system relying on this structure. On this basis, a flexibly reconfigurable energy-efficient TMA beam steering method is proposed. The energy efficiency attained is significantly improved in the dual-beam JCR operating mode conceived. In Table \ref{TabA1}, we boldly and explicitly contrast our new contributions to the relevant state-of-the-art \cite{TMA6,TMA7,TMA8,TMA9,TMA10,TMA11,TMA12,TMA13,TMA14,TMA18,EE10,EE0,TMA19,Ref1,EE13,EE14,EE15,EE22,DE1,JSTSP1}. Specifically, our contributions can be summarized as follows.

\begin{enumerate}
  \item The proposed closed-form expressions derived for TMA power loss are extended for positive symmetrically positioned pulses \cite{Close1,Ref1}, positive asymmetrically positioned pulses \cite{Close2,Ref2} and asymmetric half-cycle anti-phase periodic modulation functions \cite{JSTSP1}. The extended expressions are applicable to TMAs having a single-sideband structure (including an in-phase branch and a $1/4$-cycle delayed quadrature branch).
  \item Based on the above, the proposed closed-form expressions are applicable to both linear, planar, and volumetric (conformal) arrays.
  \item Based on \cite{TMA10,DE1} and \cite{JSTSP1}, this paper presents a quantitative study on the energy efficiency of a single-sideband TMA-based JCR system, which lays the foundations for the subsequent energy efficiency analysis and optimization.
  \item A flexibly reconfigurable energy-efficient TMA beam steering method is proposed. Compared to \cite{JSTSP1}, this solution improves energy efficiency at the cost of increasing the hardware complexity on the RF side.
  \item Finally, the implementation of a flexible and reconfigurable energy-efficient single/dual-beam TMA is discussed, which enables TMA to achieve simple single/dual-beam transformation with high energy efficiency, and the application scenario is expanded.
\end{enumerate}

\begin{figure}[!t]
\centering
\includegraphics[width=2.0in]{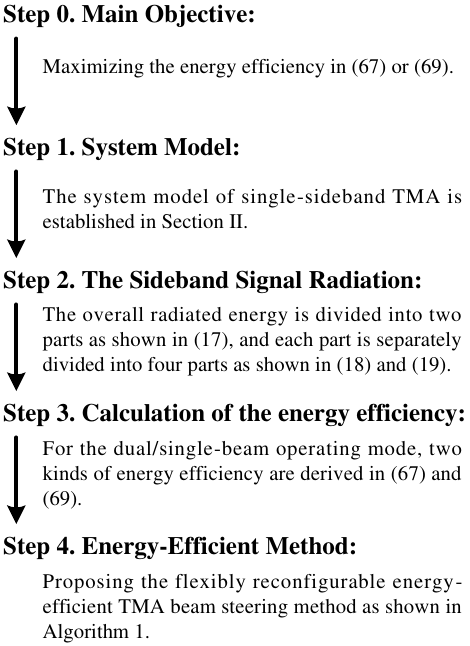}
\caption{Flow of the mathematical analysis.}
\label{Fig_A1}
\end{figure}
The remainder of this paper is organized as follows. In Section II, the preliminaries of single-sideband TMAs are introduced. In Section III, a closed-form expression is derived for the TMA power loss, and a flexibly reconfigurable (single-beam or dual-beam mode) energy-efficient TMA beam steering method is proposed. Our numerical results are presented in Section IV, and our conclusions are offered in Section V.

\section{Preliminaries}
\begin{figure}[!t]
\centering
\includegraphics[width=2.0in]{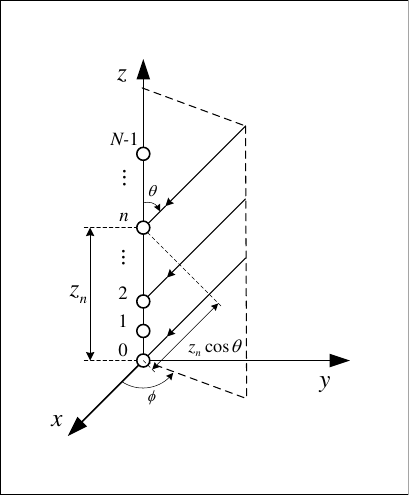}
\caption{$N$-element linear array.}
\label{Fig1}
\end{figure}

\begin{figure}[!t]
\centering
\includegraphics[width=3.2in]{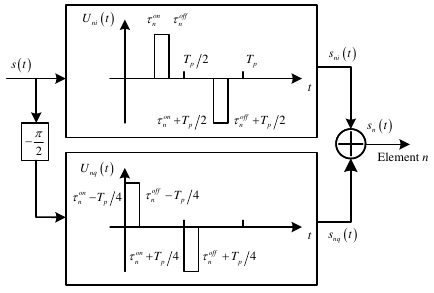}
\caption{The circuit structure before array element $n$.}
\label{Fig2}
\end{figure}

We first provide a flow of the mathematical analysis shown in Fig. \ref{Fig_A1} for easy understanding. Fig. \ref{Fig1} portrays the schematic of a TMA composed of $N$-element omnidirectional antennas in spherical spatial coordinates, arranged along the $z$-axis. The radiation field can be expressed as
\begin{equation}\label{Equ1}
  \begin{aligned}
  F(\theta,\phi,t)&=F(\theta,t)\\
  &= e^{j\omega t}\sum_{n=0}^{N-1}A_n \big(U_{ni}(t) + e^{-j\frac{\pi}{2}}U_{nq}(t)\big)e^{j\beta z_n \cos \theta},
  \end{aligned}
\end{equation}
\noindent where, $A_n = |A_n|e^{j\rm{arg}(A_n)}=a_ne^{j\varphi_n}$ is the weighted vector and $z_n$ is the coordinate of element $n$. Furthermore, $\phi$ and $\theta$ represent the azimuth and elevation angles, respectively. Still referring to (\ref{Equ1}), $\beta = 2\pi/\lambda = \omega/c$ represents the field wavenumber, where $\lambda$ is the wavelength, $\omega$ is the angular frequency of the signal carrier and $c$ represents the speed of light in vacuum. Finally, $U_{ni}(t)$ and $U_{nq}(t)$ are the periodical modulation functions of element $n$ as shown in Fig. \ref{Fig2}, where $T_p$ represents the modulation period, $\tau_{n}^{on}$ and $\tau_n^{of\!f}$ are the turn-on and turn-off times, respectively. The relationship between $U_{ni}(t)$ and $U_{nq}(t)$ satisfies
\begin{equation}\label{Equ_R5}
  U_{ni}(t) = U_{nq}(t+\frac{T_p}{4}),
\end{equation}

Fig. \ref{Fig2} shows the circuit structure before the array element $n$, i.e., the single-sideband structure TMA \cite{SSB1}, which uses symmetry to eliminate more unwanted harmonic components. We define $\alpha_{ni,k}$ and $\alpha_{nq,k}$ as the $k$th harmonic coefficients of $U_{ni}(t)$ and $U_{nq}(t)$, respectively. Then

\begin{equation}\label{Equ9}
  \left\{
  \begin{aligned}
  &\alpha_{ni,k} = \frac{1}{T_p}\int_{T_p}U_{ni}(t)e^{-j2\pi kF_pt}\rm{d}t\\
  &\alpha_{nq,k} = \frac{1}{T_p}\int_{T_p}U_{nq}(t)e^{-j2\pi kF_pt}\rm{d}t,
  \end{aligned}
  \right.
\end{equation}
\noindent where $F_p=1/T_p$ is the modulation frequency, yielding
\begin{equation}\label{Equ10}
  \begin{aligned}
  \alpha_{ni,k} =& \frac{\sin\left[\pi k \left(\xi_n^{of\!f}-\xi_n^{on}\right)\right]}{\pi k}e^{-j\pi k (\xi_n^{off}+\xi_n^{on})}\times \left(1-e^{-j\pi k}\right),
  \end{aligned}
\end{equation}
\begin{equation}\label{Equ11}
  \begin{aligned}
  \alpha_{nq,k} =& \frac{\sin\left[\pi k \left(\xi_n^{of\!f}-\xi_n^{on}\right)\right]}{\pi k} e^{-j\pi k (\xi_n^{off}+\xi_n^{on})}\\
  &\times \left(1-e^{-j\pi k}\right)e^{j\frac{\pi k}{2}}\\
  =& \alpha_{ni,k}\cdot e^{j\frac{\pi k}{2}},
  \end{aligned}
\end{equation}
\noindent where $\xi_n^{on} = F_p\tau_n^{on}$ and $\xi_n^{of\!f} = F_p\tau_n^{of\!f}$ are the normalized turn-on and turn-off times. 
\noindent From (\ref{Equ10}) and (\ref{Equ11}), the joint harmonic coefficient becomes:
\begin{equation}\label{Equ13}
  \begin{aligned}
  \alpha_{ni,k}\! +\! e^{-\!j\!\frac{\pi}{2}}\alpha_{nq,k} =& \frac{\sin\!\left[\pi k\! \left(\xi_n^{of\!f}\!-\!\xi_n^{on}\right)\right]}{\pi k}e^{-j\pi k(\xi_n^{of\!f}\!+\!\xi_n^{on})}\\
 & \times \left(1-e^{-j\pi k}\right)\left(1+e^{-j\frac{\pi}{2}}e^{j\frac{\pi k}{2}}\right).
  \end{aligned}
\end{equation}
\noindent Because $k \in \bf{Z}$, it holds that $\left(1-e^{-j\pi k}\right) = 1-\cos(k\pi) = 0$ if $k$ is even. When $k$ is odd, it holds that $\left(1+e^{-j\frac{\pi}{2}}e^{j\frac{\pi k}{2}}\right) = 1+\sin\left(\frac{\pi k}{2}\right) = 0$ if $k = -1, +3, -5, +7,\cdots$. In summary, the value of (\ref{Equ13}) is not $0$, only when we have $k = +1, -3, +5, -7, \cdots$, that is to say, the antenna pattern function
\begin{equation}\label{Equ14}
  \begin{aligned}
  F(\theta,t)=&\sum_{k=-\infty}^{\infty}\sum_{n=0}^{N-1}A_n\left(\alpha_{ni,k}+e^{-j\frac{\pi}{2}}\alpha_{nq,k}\right)\\
  &\times e^{j(\beta z_n\cos\theta + \omega t + 2\pi kF_p t)}\\
  =&\sum_{k=-\infty}^{\infty}F_k(\theta,t),
  \end{aligned}
\end{equation}
\noindent contains only components with $k = +1, -3, +5, -7, \cdots$.

The $+1$st ($k=1$) and $-3$rd ($k=-3$) components can be used for harmonic beamforming. Let $\alpha_{n,k}=\alpha_{ni,k}+ e^{-j\frac{\pi}{2}}\alpha_{nq,k}$; when $k = 1$ and $-3$, we have
\begin{equation}\label{Equ15}
  \left\{
  \begin{aligned}
  \alpha_{n,1}=& \frac{4\sin\left[\pi\left(\xi_n^{of\!f}-\xi_n^{on}\right)\right]}{\pi}e^{-j\pi\left(\xi_n^{of\!f}+\xi_n^{on}\right)} \\
  \alpha_{n,-3}=& \frac{4\sin\left[3\pi\left(\xi_n^{of\!f}-\xi_n^{on}\right)\right]}{3\pi}e^{j3\pi\left(\xi_n^{of\!f}+\xi_n^{on}\right)},
  \end{aligned}
  \right.
\end{equation}
Let the main lobe directions for the two components be $\theta_1$ and $\theta_{-3}$, respectively; then, we have
\begin{equation}\label{Equ16}
  \left\{
  \begin{aligned}
  \varphi_n - \pi\left(\xi_n^{of\!f}+\xi_n^{on}\right)+\beta z_n \cos\theta_1 =& 0 \\
  \varphi_n + 3\pi\left(\xi_n^{of\!f}+\xi_n^{on}\right)+\beta z_n \cos\theta_{-3} =& 0.
  \end{aligned}
  \right.
\end{equation}
Upon adopting the notation of
\begin{equation}\label{Equ17}
  \sin\left[\pi\left(\xi_n^{of\!f}-\xi_n^{on}\right)\right] = \sigma_n,
\end{equation}
\noindent we can obtain
\begin{equation}\label{Equ18}
  \left\{
  \begin{aligned}
  \varphi_n =& -\frac{1}{4}\beta z_n\left(3\cos\theta_1 + \cos\theta_{-3}\right) \\
  \xi_n^{on}=& \frac{\beta z_n\left(\cos\theta_1 - \cos\theta_{-3}\right)-4\arcsin\sigma_n}{8\pi}\\
  \xi_n^{of\!f}=& \frac{\beta z_n\left(\cos\theta_1 - \cos\theta_{-3}\right)+4\arcsin\sigma_n}{8\pi}.
  \end{aligned}
  \right.
\end{equation}
\noindent Harmonic beamforming of the $+1$st and $-3$rd components can be achieved in accordance with (\ref{Equ18}). The $+1$st component and the $-3$rd component (with the highest energy) can be used for radar detection and wireless communication, respectively, to realize a dual-beam JCR system. The basic principle is similar to that applied in \cite{DE1} and \cite{JSTSP1} and will not be repeated here. In addition, see \cite{Ref3,TMA10} for details on the relationship between the various harmonics of TMA and interference analysis.

\section{Sideband signal radiation calculation and optimization}
\subsection{Calculation of sideband signal radiation}\label{Section_A}
To calculate the power losses, we rewrite (\ref{Equ14}) as
\begin{equation}\label{Equ19}
  F(\theta,t) = \sum_{k=-\infty}^{\infty}|\mu_k(\theta)|e^{j\varphi_k}e^{j\left(\omega t+2\pi kF_p t\right)},
\end{equation}
\noindent where
\begin{equation}\label{Equ20}
  \mu_k(\theta) = \sum_{n=0}^{N-1}A_n\left(\alpha_{ni,k}+e^{-j\frac{\pi}{2}}\alpha_{nq,k}\right)e^{j\beta z_n\cos\theta}.
\end{equation}

The average power density over period $T_p$ is
\begin{equation}\label{Equ21}
\begin{aligned}
\tilde{P}(\theta) =& \frac{1}{T_p}\int_{0}^{T_p}|F(\theta,t)|^2\rm{d}t \\
=& \frac{1}{T_p}\int_{0}^{T_p}F(\theta,t)F^\ast(\theta,t)\rm{d}t
= \sum_{k=-\infty}^{\infty}|\mu_k(\theta)|^2.
\end{aligned}
\end{equation}
\noindent Then, the total radiated energy of the TMA is
\begin{equation}\label{Equ22}
  \begin{aligned}
  \mathscr{P} =& \int_{0}^{2\pi}\int_{0}^{\pi}\tilde{P}(\theta)\sin\theta\rm{d}\theta\rm{d}\varphi\\
  =& 2\pi\int_{0}^{\pi}\sum_{k=-\infty}^{\infty}|\mu_k(\theta)|^2\sin\theta\rm{d}\theta.
  \end{aligned}
\end{equation}
\noindent We first consider the term $|\mu_k(\theta)|^2$:
\begin{equation}\label{Equ23}
  \begin{aligned}
  |\mu_k(\theta)|^2 =& \mu_k(\theta)\mu_k^\ast(\theta) \\
  =& \sum_{n=0}^{N-1}|A_n|^2\Big(|\alpha_{ni,k}|^2 + |\alpha_{nq,k}|^2\\
  &+ e^{j\frac{\pi}{2}}\alpha_{ni,k}\alpha_{nq,k}^\ast + e^{-j\frac{\pi}{2}}\alpha_{nq,k}\alpha_{ni,k}^\ast\Big) \\
  &+\sum_{\substack{m,n = 0\\ m\neq n}}^{N-1}A_nA_m^\ast\Big( \alpha_{ni,k}\alpha_{mi,k}^\ast + \alpha_{nq,k}\alpha_{mq,k}^\ast\\
  &+e^{j\frac{\pi}{2}}\alpha_{ni,k}\alpha_{mq,k}^\ast + e^{-j\frac{\pi}{2}}\alpha_{nq,k}\alpha_{mi,k}^\ast\Big)\\
  &\times e^{j\beta(z_n-z_m)\cos\theta} \\
  =&\sum_{n=0}^{N-1}|A_n|^2\bm{\mu}_1 + \sum_{\substack{m,n=0\\m\neq n}}^{N-1}A_nA_m^\ast\bm{\mu}_2 e^{j\beta(z_n-z_m)\cos\theta}.
  \end{aligned}
\end{equation}
\noindent Upon substituting (\ref{Equ23}) into (\ref{Equ22}), we obtain
\begin{equation}\label{Equ24}
  \begin{aligned}
  \mathscr{P} =& 2\pi\sum_{k=-\infty}^{\infty}\bigg(\int_{0}^{\pi}\sum_{n=0}^{N-1}|A_n|^2\bm{\mu}_1\sin\theta\rm{d}\theta
  \\&+ \int_{0}^{\pi}\sum_{\substack{m,n=0\\m\neq n}}^{N-1}A_nA_m^\ast\bm{\mu}_2 e^{j\beta (z_n-z_m)\cos\theta}\sin\theta\rm{d}\theta \bigg) \\
  =&\mathscr{P}_1 + \mathscr{P}_2,
  \end{aligned}
\end{equation}
\noindent where
\begin{equation}\label{Equ25}
  \begin{aligned}
  \mathscr{P}_1 =&4\pi \sum_{n=0}^{N-1}|A_n|^2\sum_{k=-\infty}^{\infty}\bm{\mu}_1 \\
  =&4\pi \sum_{n=0}^{N-1}|A_n|^2\sum_{k=-\infty}^{\infty}\Big(|\alpha_{ni,k}|^2 + |\alpha_{nq,k}|^2
  \\&+ e^{j\frac{\pi}{2}}\alpha_{ni,k}\alpha_{nq,k}^\ast + e^{-j\frac{\pi}{2}}\alpha_{nq,k}\alpha_{ni,k}^\ast\Big)
  \end{aligned}
\end{equation}
\noindent and
\begin{equation}\label{Equ26}
  \begin{aligned}
  \mathscr{P}_2 =& 4\pi\sum_{\substack{m,n=0\\m\neq n}}^{N-1}A_nA_m^\ast\rm{sinc}\left[\beta\left(z_n-z_m\right)\right]\sum_{k=-\infty}^{\infty}\bm{\mu}_2 \\
  =&4\pi\sum_{\substack{m,n=0\\m\neq n}}^{N-1}A_nA_m^\ast\rm{sinc}\left[\beta\left(z_n-z_m\right)\right]\\
  &\times \sum_{k=-\infty}^{\infty}\Big( \alpha_{ni,k}\alpha_{mi,k}^\ast + \alpha_{nq,k}\alpha_{mq,k}^\ast \\
  &+e^{j\frac{\pi}{2}}\alpha_{ni,k}\alpha_{mq,k}^\ast + e^{-j\frac{\pi}{2}}\alpha_{nq,k}\alpha_{mi,k}^\ast\Big).
  \end{aligned}
\end{equation}
In the following derivation, we consider $\mathscr{P}_1$ and $\mathscr{P}_2$ separately.

\subsubsection{Calculation of $\mathscr{P}_1$}
According to (\ref{Equ25}), $\mathscr{P}_1$ consists of four important terms, $|\alpha_{ni,k}|^2$, $|\alpha_{nq,k}|^2$, $\alpha_{ni,k}\alpha_{nq,k}^\ast$ and $\alpha_{nq,k}\alpha_{ni,k}^\ast$, which are deduced separately in the following.

We first consider the term $|\alpha_{ni,k}|^2$:
\begin{equation}\label{Equ27}
  \begin{aligned}
    |\alpha_{ni,k}|^2 =& \frac{1}{\pi ^2 k^2}\left[1-\cos\left(k\pi\right)\right]
    \times \left[1-\cos 2k\pi \left(\xi_{n}^{off}-\xi_{n}^{on}\right)\right] \\
    =& \frac{1}{\pi^2 k^2} - \frac{\cos(k\pi)}{\pi^2 k^2} -\frac{\cos\left[k\cdot 2\pi\left(\xi_{n}^{off}-\xi_{n}^{on}\right)\right]}{\pi^2 k^2}
    \\&+ \frac{\cos k\left[\pi + 2\pi\left(\xi_{n}^{off}-\xi_{n}^{on}\right)\right]}{2\pi^2 k^2} \\
    & + \frac{\cos k\left[\pi - 2\pi\left(\xi_{n}^{off}-\xi_{n}^{on}\right)\right]}{2\pi^2 k^2}.
  \end{aligned}
\end{equation}
\noindent From (\ref{Equ10}), we have
\begin{equation}\label{Equ28}
  \left\{\begin{aligned}
  & |\alpha_{ni,k}|^2 = |\alpha_{ni,(-k)}|^2 \\
  & \alpha_{ni,0} = 0
  \end{aligned}\right..
\end{equation}
\noindent According to Appendix \ref{AppendixA} we also have:
\begin{equation}\label{Equ29}
  \begin{aligned}
    \sum_{k= -\infty}^{\infty}|\alpha_{ni,k}|^2=& 2\sum_{k=1}^{\infty}|\alpha_{ni,k}|^2
    =& 2\tau_n,
  \end{aligned}
\end{equation}
\noindent where $\tau_n = \xi_{n}^{off}-\xi_{n}^{on}$ represents the normalized turned-on duration of element $n$.

Secondly, we consider the term $|\alpha_{nq,k}|^2$ from (\ref{Equ11}):
\begin{equation}\label{Equ31}
  |\alpha_{nq,k}|^2=|\alpha_{ni,k}e^{j\frac{\pi k}{2}}|^2 = |\alpha_{ni,k}|^2;
\end{equation}
\noindent then, we have:
\begin{equation}\label{Equ32}
  \sum_{k=-\infty}^{\infty}|\alpha_{nq,k}|^2 = 2\tau_n.
\end{equation}

Thirdly, we consider the term $\alpha_{ni,k}\alpha_{nq,k}^\ast$ from (\ref{Equ10}) and (\ref{Equ11}):
\begin{equation}\label{Equ33}
  \alpha_{ni,k}\alpha_{nq,k}^\ast = |\alpha_{ni,k}|^2\left(\cos\frac{\pi k}{2}-j\sin\frac{\pi k}{2}\right),
\end{equation}
\noindent where
\begin{equation}\label{Equ34}
  \left\{\begin{aligned}
  |\alpha_{ni,-k}|^2 =& |\alpha_{ni,k}|^2 \\
  \cos\frac{\pi(-k)}{2} =& \cos\frac{\pi k}{2} \\
  \sin\frac{\pi (-k)}{2} =& -\sin\frac{\pi k}{2}.
  \end{aligned}\right.
\end{equation}
\noindent Then, from Appendix \ref{AppendixB} we have
\begin{equation}\label{Equ35}
  \begin{aligned}
  \sum_{k=-\infty}^{\infty}\alpha_{ni,k}\alpha_{nq,k}^{\ast} =& \sum_{k = -\infty}^{\infty}|\alpha_{ni,k}|^2e^{-j\frac{\pi k}{2}}
  \\=& 2\sum_{k=1}^{\infty}|\alpha_{ni,k}|^2\cos\frac{\pi k}{2}
  =0.
  \end{aligned}
\end{equation}

Fourthly, we consider the term $\alpha_{nq,k}\alpha_{ni,k}^{\ast}$ and we obtain the same result:
\begin{equation}\label{Equ47}
  \sum_{k=-\infty}^{\infty}\alpha_{nq,k}\alpha_{ni,k}^{\ast} = 0.
\end{equation}
\noindent Substituting (\ref{Equ29}), (\ref{Equ32}), (\ref{Equ35}) and (\ref{Equ47}) into (\ref{Equ25}) yields
\begin{equation}\label{Equ48}
  \mathscr{P}_1 = 16\pi \sum_{n=0}^{N-1}\left(|A_n|^2\tau_n\right).
\end{equation}

\subsubsection{Calculation of $\mathscr{P}_2$}
According to (\ref{Equ26}), $\mathscr{P}_2$ consists of four important terms, $\alpha_{ni,k}\alpha_{mi,k}^{\ast}$, $\alpha_{nq,k}\alpha_{mq,k}^{\ast}$, $\alpha_{ni,k}\alpha_{mq,k}^{\ast}$ and
$\alpha_{nq,k}\alpha_{mi,k}^{\ast}$, which are deduced separately in the following.

\begin{figure}[!t]
\centering
\includegraphics[width=2.5in]{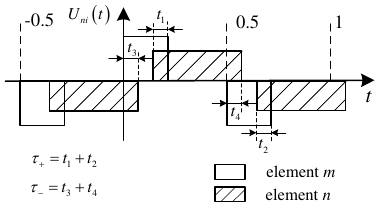}
\caption{Illustration of $\tau_+$ and $\tau_-$\cite{JSTSP1}.}
\label{Fig3}
\end{figure}

We first consider the term $\alpha_{ni,k}\alpha_{mi,k}^{\ast}$ from (\ref{Equ10}):
\begin{equation}\label{Equ49}
  \begin{aligned}
    \alpha_{ni,k}\alpha_{mi,k}^{\ast} =& \frac{1-\cos(\pi k)}{\pi^2 k^2}\big[\cos\left(\pi k\left(\tau_n-\tau_m\right)\right)
    \\&- \cos\left(\pi k\left(\tau_n+\tau_m\right)\right)\big] \\
    &\times \big[\cos\left(\pi k\left(\xi_{m}^{off}-\xi_{n}^{off}+\xi_{m}^{on}-\xi_{n}^{on}\right)\right) \\
    &+j\sin\left(\pi k\left(\xi_{m}^{off}-\xi_{n}^{off}+\xi_{m}^{on}-\xi_{n}^{on}\right)\right)\big].
  \end{aligned}
\end{equation}
Upon considering the parity of (\ref{Equ49}), we have
\begin{equation}\label{Equ50}
  \begin{aligned}
  \sum_{k=-\infty}^{\infty}\alpha_{ni,k}\alpha_{mi,k}^{\ast} =& 2\sum_{k=1}^{\infty}\frac{1-\cos(\pi k)}{\pi^2 k^2}\big(\cos\left[\pi k\left(\tau_n-\tau_m\right)\right]
  \\&-\cos\left[\pi k\left(\tau_n+\tau_m\right)\right]\big)\\
  &\times \cos\big[\pi k(\xi_{m}^{off}
  -\xi_{n}^{off}+\xi_{m}^{on}-\xi_{n}^{on})\big] \\
  =& \sum_{k=1}^{\infty}\frac{1-\cos(\pi k)}{\pi^2 k^2}
  \\&\times\big(\cos\left[2\pi k\left(\xi_{m}^{on}-\xi_{n}^{on}\right)\right]
  \\&+ \cos\left[2\pi k\left(\xi_{m}^{off}-\xi_{n}^{off}\right)\right] \\
  &-\cos\left[2\pi k\left(\xi_{m}^{off}-\xi_{n}^{on}\right)\right]
  \\&-\cos\left[2\pi k\left(\xi_{m}^{on}-\xi_{n}^{off}\right)\right]\big).
  \end{aligned}
\end{equation}
\noindent When $k$ is even, $\alpha_{ni,k}\alpha_{mi,k}^{\ast}=0$; then,
\begin{equation}\label{Equ51}
  \begin{aligned}
  \sum_{k=-\infty}^{\infty}\alpha_{ni,k}\alpha_{mi,k}^{\ast}=&\frac{2}{\pi^2}\sum_{k=1}^{\infty}\frac{1}{(2k-1)^2}
  \\&\times\big(\cos\left[2\pi\left(2k-1\right)\left(\xi_{m}^{off}-\xi_{n}^{off}\right)\right] \\
  &-\cos\left[2\pi\left(2k-1\right)\left(\xi_{m}^{off}-\xi_{n}^{on}\right)\right]
  \\&-\cos\left[2\pi\left(2k-1\right)\left(\xi_{m}^{on}-\xi_{n}^{off}\right)\right] \\
  &+\cos\left[2\pi\left(2k-1\right)\left(\xi_{m}^{on}-\xi_{n}^{on}\right)\right]\big),
  \end{aligned}
\end{equation}
\noindent which has the same form as equation (39) in \cite{JSTSP1}. Thus, we have the result that \cite{JSTSP1}
\begin{equation}\label{Equ52}
  \sum_{k=-\infty}^{\infty}\alpha_{ni,k}\alpha_{mi,k}^{\ast} = \tau_+ - \tau_-,
\end{equation}
\noindent where $\tau_+$ and $\tau_-$ are the sums of the same- and different-phase overlapping parts, respectively, of the modulation function between the elements $m$ ($U_{mi}(t)$) and $n$ ($U_{ni}(t)$) in one cycle (normalization with respect to the period), as illustrated in Fig. \ref{Fig3}.

Secondly, we consider the term $\alpha_{nq,k}\alpha_{mq,k}^{\ast}$ and we get the same result that
\begin{equation}\label{Equ53}
  \sum_{k=-\infty}^{\infty}\alpha_{nq,k}\alpha_{mq,k}^{\ast} = \tau_+ - \tau_-.
\end{equation}

Thirdly, we consider the term $\alpha_{ni,k}\alpha_{mq,k}^\ast$ from (\ref{Equ10}) and (\ref{Equ11}):
\begin{equation}\label{Equ54}
  \begin{aligned}
    \alpha_{ni,k}\alpha_{mq,k}^{\ast} =& \alpha_{ni,k}\alpha_{mi,k}^{\ast}e^{-j\frac{\pi k}{2}}
    \\=& \frac{1-\cos(\pi k)}{\pi^2 k^2}\big(\cos\left[\pi k\left(\tau_n-\tau_m\right)\right]
    \\&-\cos\left[\pi k\left(\tau_n+\tau_m\right)\right]\big)e^{j\Psi_k},
  \end{aligned}
\end{equation}
\noindent where we have
\begin{equation}\label{Equ55}
  \Psi_k =k\left(\pi\left(\xi_{m}^{off}-\xi_{n}^{off}+\xi_{m}^{on}-\xi_{n}^{on}\right)-\frac{\pi}{2}\right)
\end{equation}
\noindent and
\begin{equation}\label{Equ56}
  \begin{aligned}
    e^{j\Psi_k} = \cos\Psi_k + j\sin\Psi_k.
  \end{aligned}
\end{equation}
\noindent Upon considering the parity of (\ref{Equ54}), we have
\begin{equation}\label{Equ57}
  \begin{aligned}
    \sum_{k=-\infty}^{\infty}\alpha_{ni,k}\alpha_{mq,k}^{\ast} =&2\sum_{k=-\infty}^{\infty}\frac{1-\cos(\pi k)}{\pi^2 k^2}\big(\cos\left[\pi k\left(\tau_n-\tau_m\right)\right]
    \\&-\cos\left[\pi k\left(\tau_n+\tau_m\right)\right]\big)\cos\Psi_k \\
    =& \frac{1}{\pi^2}\sum_{k=1}^{\infty}\frac{1-\cos(\pi k)}{k^2}
    \\&\times \Bigg( \cos\left(2\pi k\left(\xi_{m}^{on}-\xi_{n}^{on}\right)-\frac{\pi k}{2}\right) \\
    &+\cos\left(2\pi k\left(\xi_{m}^{off}-\xi_{n}^{off}\right)-\frac{\pi k}{2}\right)
    \\&-\cos\left(2\pi k \left(\xi_{m}^{off}-\xi_{n}^{on}\right)-\frac{\pi k}{2}\right) \\
    &- \cos\left(2\pi k\left(\xi_{m}^{on}-\xi_{n}^{off}\right)-\frac{\pi k}{2}\right)\Bigg).
  \end{aligned}
\end{equation}
\noindent We consider one of the terms in (\ref{Equ57}), $\cos\left(2\pi k\left(\xi_{m}^{on}-\xi_{n}^{on}\right)-\frac{\pi k}{2}\right)$, for illustration:
\begin{equation}\label{Equ58}
  \begin{aligned}
    &\cos\left(2\pi k\left(\xi_{m}^{on}-\xi_{n}^{on}\right)-\frac{\pi k}{2}\right)
     \\=& \cos\left[2\pi k\left(\xi_{m}^{on}-\xi_{n}^{on}\right)\right]\cos\left(\frac{\pi k}{2}\right)
    \\&+ \sin\left[2\pi k\left(\xi_{m}^{on}-\xi_{n}^{on}\right)\right]\sin\left(\frac{\pi k}{2}\right).
  \end{aligned}
\end{equation}
\noindent When $k$ is even, $\alpha_{ni,k}\alpha_{mq,k}^{\ast} = 0$. When $k$ is odd, we have $\cos\left(\frac{\pi k}{2}\right) =0$ and $\sin\left(\frac{\pi k}{2}\right) = \pm 1$; then,
\begin{equation}\label{Equ59}
  \begin{aligned}
    \sum_{k=-\infty}^{\infty}\alpha_{ni,k}\alpha_{mq,k}^{\ast} =& \frac{2}{\pi^2}\sum_{k=0}^{\infty}\frac{(-1)^k}{(2k+1)^2}
    \\&\times \Big( \sin\left(\left(2k+1\right)2\pi\left(\xi_{m}^{on}-\xi_{n}^{on}\right)\right) \\
    &+\sin\left(\left(2k+1\right)2\pi\left(\xi_{m}^{off}-\xi_{n}^{off}\right)\right)
    \\&-\sin\left(\left(2k+1\right)2\pi\left(\xi_{m}^{off}-\xi_{n}^{on}\right)\right) \\
    &-\sin\left(\left(2k+1\right)2\pi\left(\xi_{m}^{on}-\xi_{n}^{off}\right)\right)\Big).
  \end{aligned}
\end{equation}

\begin{figure*}[!t]
\setlength{\abovecaptionskip}{0pt}
\setlength{\belowcaptionskip}{-15pt}
\centering
\includegraphics[width=5.3in]{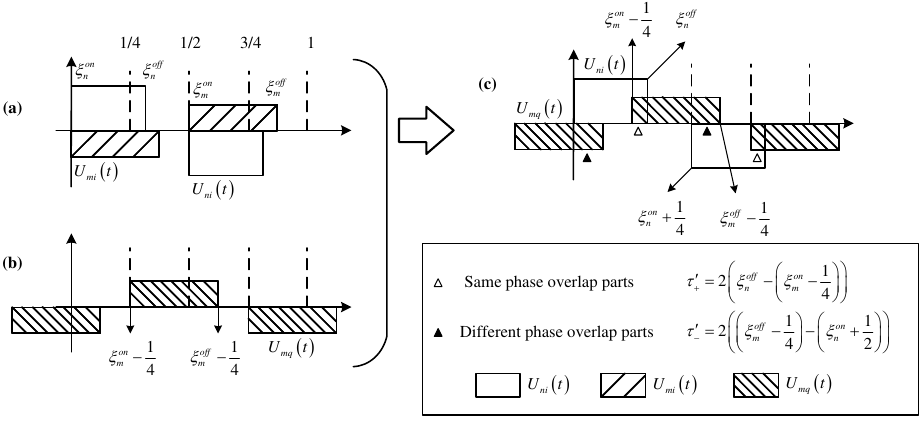}
\caption{Illustration of $\tau'_+$ and $\tau'_-$ (for the example of case 1f).}
\label{Fig5}
\end{figure*}

\noindent According to Appendix \ref{AppendixC} we get
\begin{equation}\label{Equ64}
  \begin{aligned}
    \sum_{k=-\infty}^{\infty} \alpha_{ni,k}\alpha_{mq,k}^{\ast}
    =& \tau'_+ - \tau'_-,
  \end{aligned}
\end{equation}

\noindent where $\tau'_+$ and $\tau'_-$ are the sums of the same- and different-phase overlapping parts, respectively, of the modulation function between elements $n$ ($U_{ni}(t)$) and $m$ ($U_{mq}(t)$) in one cycle (normalized with respect to the period), as illustrated in Fig.~\ref{Fig5}.

In Fig.~\ref{Fig5}, case 1f (in Fig.~\ref{Fig4}) is taken as an example to illustrate the definitions of $\tau'_{+}$ and $\tau'_-$. In Fig. \ref{Fig5}, part (a) shows the modulation functions $U_{ni}(t)$ and $U_{mi}(t)$, part (b) portrays the modulation function $U_{mq}(t)$, and part (c) represents the position relationship between $U_{ni}(t)$ and $U_{mq}(t)$. The definitions of $\tau'_+$ and $\tau'_-$ are also illustrated.

Fourthly, we consider the term $\alpha_{nq,k}\alpha_{mi,k}^{\ast}$. Similar to the derivation process of (\ref{Equ54}) to (\ref{Equ59}), we obtain
\begin{equation}\label{Equ65}
  \sum_{k=-\infty}^{\infty}\alpha_{nq,k}\alpha_{mi,k}^{\ast} = -\sum_{k=-\infty}^{\infty}\alpha_{ni,k}\alpha_{mq,k}^{\ast}.
\end{equation}
\noindent Substituting (\ref{Equ52}), (\ref{Equ53}), (\ref{Equ64}) and (\ref{Equ65}) into (\ref{Equ26}) yields
\begin{equation}\label{Equ66}
  \begin{aligned}
  \mathscr{P}_2 =& 8\pi\sum_{\substack{m,n=0\\m\neq n}}^{N-1}A_nA_m^{\ast}{\rm sinc}\left[\beta\left(z_n-z_m\right)\right]
  \\&\times \left[\left(\tau_+-\tau_-\right)+j\left(\tau'_+-\tau'_-\right)\right].
  \end{aligned}
\end{equation}

For convenience, we adopt the following definitions:
\begin{equation}\label{Equ67}
    \sum_{k=-\infty}^{\infty}\alpha_{ni,k}\alpha_{mi,k}^{\ast} =\sum_{k=-\infty}^{\infty}\alpha_{nq,k}\alpha_{mq,k}^{\ast} = \tau_{nm}.
\end{equation}
\noindent From (\ref{Equ51}), we have
\begin{equation}\label{Equ68}
\sum_{k=-\infty}^{\infty}\alpha_{ni,k}\alpha_{mi,k}^{\ast} = \sum_{k=-\infty}^{\infty}\alpha_{mi,k}\alpha_{ni,k}^{\ast}.
\end{equation}
\noindent Similarly,
\begin{equation}\label{Equ69}
\sum_{k=-\infty}^{\infty}\alpha_{nq,k}\alpha_{mq,k}^{\ast} = \sum_{k=-\infty}^{\infty}\alpha_{mq,k}\alpha_{nq,k}^{\ast}.
\end{equation}
\noindent We define that
\begin{equation}\label{Equ70}
    \sum_{k=-\infty}^{\infty}\alpha_{mi,k}\alpha_{ni,k}^{\ast} =\sum_{k=-\infty}^{\infty}\alpha_{mq,k}\alpha_{nq,k}^{\ast} = \tau_{mn}.
\end{equation}
Then it may be seen that
\begin{equation}\label{Equ71}
  \tau_{nm} = \tau_{mn} = \tau_+-\tau_-.
\end{equation}

In addition, we define that
\begin{equation}\label{Equ72}
  \sum_{k=-\infty}^{\infty}\alpha_{ni,k}\alpha_{mq,k}^{\ast} = -\sum_{k=-\infty}^{\infty}\alpha_{nq,k}\alpha_{mi,k}^{\ast} = \tau'_{nm}.
\end{equation}
\noindent From (\ref{Equ59}), we have
\begin{equation}\label{Equ73}
  \sum_{k=-\infty}^{\infty}\alpha_{ni,k}\alpha_{mq,k}^{\ast}=
  -\sum_{k=-\infty}^{\infty}\alpha_{mi,k}\alpha_{nq,k}^{\ast}.
\end{equation}
\noindent Similarly,
\begin{equation}\label{Equ74}
  \sum_{k=-\infty}^{\infty}\alpha_{nq,k}\alpha_{mi,k}^{\ast}=
  -\sum_{k=-\infty}^{\infty}\alpha_{mq,k}\alpha_{ni,k}^{\ast}.
\end{equation}
\noindent We define that
\begin{equation}\label{Equ75}
  \sum_{k=-\infty}^{\infty}\alpha_{mi,k}\alpha_{nq,k}^{\ast} = -\sum_{k=-\infty}^{\infty}\alpha_{mq,k}\alpha_{ni,k}^{\ast} = \tau'_{mn}.
\end{equation}
\noindent Then it is readily seen that
\begin{equation}\label{Equ76}
  \tau'_{nm} = -\tau'_{mn} = \tau'_+-\tau'_-.
\end{equation}
Then,
\begin{equation}\label{Equ77}
  \mathscr{P}_2 = 8\pi\sum_{\substack{m,n=0\\m\neq n}}^{N-1}A_nA_m^{\ast}{\rm sinc}\left[\beta\left(z_n-z_m\right)\right] (\tau_{nm}+j\tau'_{nm}).
\end{equation}

Upon selecting $m$ and $n$ arbitrarily, we perform the following calculation:
\begin{equation}\label{Equ78}
  \begin{aligned}
  &A_nA_m^{\ast}{\rm sinc}\left[\beta\left(z_n-z_m\right)\right](\tau_{nm}+j\tau'_{nm})
  \\&+ A_mA_n^{\ast}{\rm sinc}\left[\beta\left(z_m-z_n\right)\right](\tau_{mn}+j\tau'_{mn}) \\
  =&{\rm sinc}\left[\beta\left(z_n-z_m\right)\right]\big[\tau_{nm}(A_nA_m^{\ast}+A_mA_n^{\ast})
  \\&+j\tau'_{nm}(A_nA_m^{\ast}-A_mA_n^{\ast})\big] \\
  =&2{\rm sinc}\left[\beta\left(z_n-z_m\right)\right]\big[\tau_{nm}{\rm Re(A_nA_m^{\ast})}
  \\&+\tau'_{nm}{\rm Im(A_nA_m^{\ast})}\big],
  \end{aligned}
\end{equation}
\noindent where ${\rm Re}(x)$ denotes the real part of $x$ and ${\rm Im}(x)$ denotes the imaginary part of $x$. Then, (\ref{Equ77}) can be written as
\begin{equation}\label{Equ79}
  \begin{aligned}
    \mathscr{P}_2 =& 16\pi\!\! \sum_{m,n\in \Theta_{mn}}\!\!{\rm sinc}\left[\beta\left(z_n-z_m\right)\right]\big[\tau_{nm}{\rm Re}(A_nA_m^{\ast})
    \\&+\tau'_{nm}{\rm Im}(A_nA_m^{\ast})\big],
  \end{aligned}
\end{equation}
\noindent where $\Theta_{mn}$ is the index set corresponding to all non-repeated $(m,n)$ pairs with $m \neq n$. For example, if $N=3$, then $\Theta_{mn} = \{(0,1),(0,2),(1,2)\}$. Upon substituting (\ref{Equ48}) and (\ref{Equ79}) into (\ref{Equ24}), we obtain the total radiated energy of the TMA:
\begin{equation}\label{Equ80}
  \begin{aligned}
    \mathscr{P} =& 16\pi\sum_{n=0}^{N-1}\left(|A_n|^2\tau_n\right) + 16\pi\!\!\! \sum_{m,n\in \Theta_{mn}}\!\!\!{\rm sinc}\left[\beta\left(z_n-z_m\right)\right]\\
    &\times \big[\tau_{nm}{\rm Re}(A_nA_m^{\ast})+\tau'_{nm}{\rm Im}(A_nA_m^{\ast})\big].
  \end{aligned}
\end{equation}

\subsubsection{Energy of useful components}
Depending on the operating mode, the useful components are defined differently. If the dual-beam operating mode is adopted, both the $+1$st and $-3$rd harmonic components are useful components. If the single-beam operating mode is adopted, only the $+1$st harmonic component is useful. Here, the amounts of energy in the $+1$st and $-3$rd harmonic components are calculated.

We denote the energy of the $+1$st harmonic component by $\mathscr{P}_{{\rm useful}+1}$ and define it as
\begin{equation}\label{Equ81}
  \mathscr{P}_{{\rm useful}+1} = \mathscr{P}_1|_{k=1} + \mathscr{P}_2|_{k=1},
\end{equation}
\noindent where we have
\begin{equation}\label{Equ82}
  \begin{aligned}
    \mathscr{P}_1|_{k=1} =& 4\pi \sum_{n=0}^{N-1}|A_n|^2\big(|\alpha_{ni,1}|^2+|\alpha_{nq,1}|^2
    \\&+e^{j\frac{\pi}{2}}\alpha_{ni,1}\alpha_{nq,1}^{\ast}+e^{-j\frac{\pi}{2}}\alpha_{nq,1}\alpha_{ni,1}^{\ast}\big),
  \end{aligned}
\end{equation}
\begin{equation}\label{Equ83}
  \begin{aligned}
    \mathscr{P}_2|_{k=1} =& 4\pi\sum_{\substack{m,n=0\\m\neq n}}^{N-1}A_nA_m^{\ast}{\rm sinc}\left[\beta\left(z_n-z_m\right)\right]
    \\&\times\big[\alpha_{ni,1}\alpha_{mi,1}^{\ast}+\alpha_{nq,1}\alpha_{mq,1}^{\ast}
    \\&+e^{j\frac{\pi}{2}}\alpha_{ni,1}\alpha_{mq,1}^{\ast}+e^{-j\frac{\pi}{2}}\alpha_{nq,1}\alpha_{mq,1}^{\ast}\big].
  \end{aligned}
\end{equation}

\noindent After some simple derivation, we have
\begin{equation}\label{Equ84}
\begin{aligned}
  \mathscr{P}_{{\rm useful}+1} =& \frac{32}{\pi}\sum_{n=0}^{N-1}|A_n|^2\left[1-\cos\left(2\pi\tau_n\right)\right]
  \\&+\frac{128}{\pi}\sum_{m,n\in \Theta_{mn}}{\rm sinc}\left[\beta\left(z_n-z_m\right)\right]\\
  &\times\sin(\pi\tau_n)\sin(\pi\tau_m)
  \\&\times {\rm Re}\left(A_nA_m^{\ast}e^{j\pi(\xi_m^{off}-\xi_n^{off}+\xi_m^{on}-\xi_n^{on})}\right).
\end{aligned}
\end{equation}

Similarly, we denote the energy of the $-3$rd harmonic component by $\mathscr{P}_{{\rm useful}-3}$ and define it as
\begin{equation}\label{Equ85}
  \mathscr{P}_{{\rm useful}-3} = \mathscr{P}_1|_{k=-3} + \mathscr{P}_2|_{k=-3},
\end{equation}
\noindent where
\begin{equation}\label{Equ86}
\begin{aligned}
  \mathscr{P}_1|_{k=-3}=&4\pi\sum_{n=0}^{N-1}|A_n|^2\big(|\alpha_{ni,-3}|^2+|\alpha_{nq,-3}|^2
  \\&+e^{j\frac{\pi}{2}}\alpha_{ni,-3}\alpha_{nq,-3}^{\ast} +e^{-j\frac{\pi}{2}}\alpha_{nq,-3}\alpha_{ni,-3}^{\ast}\big),
\end{aligned}
\end{equation}
\begin{equation}\label{Equ87}
  \begin{aligned}
    \mathscr{P}_2|_{k=-3}=&4\pi \sum_{\substack{m,n=0\\
    m\neq n}}^{N-1}\big(\alpha_{ni,-3}\alpha_{mi,-3}^{\ast}+\alpha_{nq,-3}\alpha_{mq,-3}^{\ast}
    \\&+e^{j\frac{\pi}{2}}\alpha_{ni,-3}\alpha_{mq,-3}^{\ast} + e^{-j\frac{\pi}{2}}\alpha_{nq,-3}\alpha_{mi,-3}^{\ast}\big).
  \end{aligned}
\end{equation}
\noindent Then, we have
\begin{equation}\label{Equ88}
  \begin{aligned}
    \mathscr{P}_{{\rm useful}-3} =& \frac{32}{9\pi}\sum_{n=0}^{N-1}|A_n|^2\left[1-\cos\left(6\pi \tau_n\right)\right]
    \\&+\frac{128}{9\pi}\sum_{m,n\in \Theta_{mn}}{\rm sinc}\left[\beta\left(z_n-z_m\right)\right] \\
    &\times \sin(3\pi \tau_n)\sin(3\pi \tau_m)
    \\&\times {\rm Re}\left(A_nA_m^{\ast}e^{j3\pi(\xi_n^{off}-\xi_m^{off}+\xi_n^{on}-\xi_m^{on})}\right).
  \end{aligned}
\end{equation}

\subsubsection{Calculation of the power loss}
Similarly, depending on the operating mode, the power loss is defined differently.

In the dual-beam operating mode, we define the power loss $\mathscr{P}_{\rm dual-loss}$ as
\begin{equation}\label{Equ89}
  \mathscr{P}_{\rm dual-loss} = \mathscr{P} - \mathscr{P}_{{\rm useful}+1} -\mathscr{P}_{{\rm useful}-3},
\end{equation}
\noindent and the corresponding energy efficiency $\eta_{\rm dual}$ as
\begin{equation}\label{Equ90}
  \eta_{\rm dual} = \frac{\mathscr{P}_{{\rm useful}+1} +\mathscr{P}_{{\rm useful}-3}}{\mathscr{P}}.
\end{equation}

In the single-beam operating mode, we define the power loss $\mathscr{P}_{\rm single-loss}$ as
\begin{equation}\label{Equ91}
  \mathscr{P}_{\rm single-loss} = \mathscr{P} - \mathscr{P}_{{\rm useful}+1},
\end{equation}
and the corresponding energy efficiency $\eta_{\rm single}$ as
\begin{equation}\label{Equ92}
  \eta_{\rm single} = \frac{\mathscr{P}_{{\rm useful}+1}}{\mathscr{P}}.
\end{equation}

\subsubsection{Extension to planar or volumetric (conformal) antenna arrays}
For an $S$-element planar or volumetric (conformal) antenna array, the position vector of element $s$ is $\vec{r}_s=[x_s,y_s,z_s]$, where $s=1,2,\ldots,S$. We define the signal incident direction vector $\vec{a}_r = [\sin\theta\cos\phi, \sin\theta\sin\phi, \cos\theta]$, then (\ref{Equ1}) changes as
\begin{equation}\label{Equ_R1}
\begin{aligned}
  F(\theta,\phi,t) =& e^{j\omega t}\sum_{s=1}^{S}A_sU_s(t)e^{j\beta \vec{r}_s \cdot\vec{a}_r}\\=&e^{j\omega t}\sum_{s=1}^{S}A_sU_s(t)e^{j\beta (x_s\sin\theta\cos\phi + y_s\sin\theta\sin\phi + z_s\cos\theta)}.
\end{aligned}
\end{equation}
\noindent Review the derivation process in Section \ref{Section_A}, the main difference is between (\ref{Equ26}) and (\ref{Equ_R2}), and the integral $P'(\theta,\phi)$ simplifies to (\ref{Equ_R3}) as shown at the top of the next page (refer to \cite{Ref1,Ref2}), where $R$ represents the Euclidean distance between point $\vec{r}_s=[x_s,y_s,z_s]$ and $\vec{r}_{s'}=[x_{s'},y_{s'},z_{s'}]$
\begin{figure*}
\begin{equation}\label{Equ_R2}
\begin{aligned}
  \mathscr{P}_2 =& \sum_{\substack{s,s'=1\\s\neq s'}}^{S}A_sA_{s'}^{\ast}\int_{0}^{2\pi}\int_{0}^{\pi}e^{j\beta[(x_s-x_{s'})\sin\theta\cos\phi+(y_s-y_{s'})\sin\theta\sin\phi+(z_s-z_{s'})\cos\theta]}\sin\theta {\rm d}\theta{\rm d}\phi \\
  =& \sum_{\substack{s,s'=1\\s\neq s'}}^{S}A_sA_{s'}^{\ast}P'(\theta,\phi)\sum_{k=-\infty}^{\infty}\bm{\mu}_2
\end{aligned}
\end{equation}
\end{figure*}
\vspace{-10pt}
\begin{figure*}
\begin{equation}\label{Equ_R3}
  P'(\theta,\phi)=\int_{0}^{2\pi}\int_{0}^{\pi}e^{j\beta[(x_s-x_{s'})\sin\theta\cos\phi+(y_s-y_{s'})\sin\theta\sin\phi+(z_s-z_{s'})\cos\theta]}\sin\theta {\rm d}\theta{\rm d}\phi = 4\pi \frac{\sin(\beta R)}{\beta R}
\end{equation}
\hrulefill
\end{figure*}

\begin{equation}\label{Equ_R4}
  R = \sqrt{(x_s-x_{s'})^2+(y_s-y_{s'})^2+(z_s-z_{s'})^2}.
\end{equation}
\noindent Other derivations include that $\sum_{k=-\infty}^{\infty}{\bm \mu}_1$ and $\sum_{k=-\infty}^{\infty}{\bm \mu}_2$ remain basically unchanged, so we can simply replace ${\rm sinc}[\beta(z_n-z_m)]$ with $\frac{\sin(\beta R)}{\beta R}$ in (\ref{Equ80}), (\ref{Equ84}), and (\ref{Equ88}) to obtain the corresponding closed-form energy-efficiency expression for a planar or volumetric (conformal) antenna array.

\subsubsection{Toy example}
\begin{figure}[!t]
\centering
\setlength{\abovecaptionskip}{0pt}
\setlength{\belowcaptionskip}{-15pt}
\includegraphics[width=2.9in]{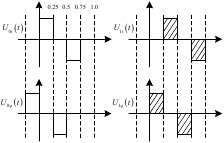}
\caption{The illustration of modulation functions $U_{0i}(t),\ U_{0q}(t)$ (for element $0$) and $U_{1i}(t),\ U_{1q}(t)$ (for element $1$) in a two-element TMA.}
\label{Fig_A2}
\end{figure}

Here shows a very simple low-dimensional toy example (TMA containing two elements with the element space $\lambda /4$) introducing the basics. The modulation functions $U_{0i}(t), U_{0q}(t)$ (for element $0$) and $U_{1i}(t), U_{1q}(t)$ (for element $1$) are illustrated in Fig. \ref{Fig_A2}. Then we have $\xi_0^{on}=0,\ \xi_0^{off}=0.25,\ \xi_1^{on}=0.25,\ \xi_1^{off}=0.5$. We assume that the weights are $A_0 = A_1 = 1$. In (\ref{Equ80}), it's obvious that $|A_0|^2 = |A_1|^2 = 1$, $\tau_0 = \tau_1 = 0.25$, $\beta (z_0 - z_1) = -\pi /2$; $\tau_{01} = 0$ (where, $\tau_+ = 0$ and $\tau_- = 0$); $\tau'_{01} = 0.5$ (where, $\tau'_+ = 0.5$ and $\tau'_- = 0$); ${\rm Re}(A_0A_1^\ast) = 1$ and ${\rm Im}(A_0A_1^\ast) = 0$. Then we have $\mathscr{P} = 25.13$. Similarly, from (\ref{Equ84}) and (\ref{Equ88}) we have $\mathscr{P}_{{\rm useful}+1} = 20.37$ and $\mathscr{P}_{{\rm useful}-3} = 2.26$. Finally, we get the energy efficiency $\eta_{\rm dual} = 90.05\%$ and $\eta_{\rm single} = 81.06\%$ from (\ref{Equ90}) and (\ref{Equ92}), respectively.

\subsection{Power loss optimization}
Based on the previous derivation results, we propose a flexibly reconfigurable (dual-beam or single-beam mode) energy-efficient TMA beam steering method. As a commonly used global optimization approach, DE is chosen to optimize the power efficiency of the TMA \cite{DE_algorithm}. The parameters $\Sigma = [\sigma_1,\ldots,\sigma_N]$ are optimized by minimizing the following cost function:
\begin{equation}\label{Equ93}
  f(\Sigma) = W_{\rm SLL}\Psi^{\rm SLL}(\Sigma)+W_{\rm Loss}P^{\rm Loss}_G(\Sigma),
\end{equation}
\noindent where $W_{\rm SLL}$ and $W_{\rm Loss}$ are real weight coefficients of $\Psi^{\rm SLL}(\Sigma)$ and $P^{\rm Loss}_G(\Sigma)$, respectively, and
\begin{equation}\label{Equ94}
  \begin{aligned}
    \Psi^{\rm SLL}(\Sigma) = \frac{H(SLL_G(\Sigma)-SLL^{\rm ref})|SLL_G(\Sigma)-SLL^{\rm ref}|^2}{|SLL^{\rm ref}|^2},
  \end{aligned}
\end{equation}
\begin{equation}\label{Equ95}
  P_G^{\rm Loss}(\Sigma)=\left\{
  \begin{aligned}
    &1-\eta_{\rm dual},\ \text{\rm Dual-beam\ mode}\\
    &1-\eta_{\rm single},\ \text{\rm Single-beam\ mode}.
  \end{aligned}\right.
\end{equation}
\noindent In (\ref{Equ94}), $H(\cdot)$ is the Heaviside step function, and $G$ is the iteration index. Briefly, (\ref{Equ94}) is a constraint on the antenna array pattern for the $+1$st harmonic and defines the distance between the current sidelobe level $SLL_G(\Sigma)$ and the desired (or reference) sidelobe level $SLL^{\rm ref}$. Furthermore, (\ref{Equ95}) defines the power loss of the TMA depending on the operating mode. The complete process is summarized in Algorithm 1. Note that in the single-beam operating mode, the direction of the -3rd harmonic can be arbitrarily set. We calculated the number of multiplications and additions required for an iteration, resulting in a computational complexity order of approximately $O(N^2N\!P)$. This is the same as in \cite{JSTSP1}, because the associated closed-form expression is similar, even though the modulation mode is changed.

In this paper, we mainly consider scenarios where the radar and communication functions are activated simultaneously, even though in practice they do not always overlap on the timeline \cite{ISAC1}. Considering the asymmetry of the two service frequencies, the resource utilization efficiency may be further improved, which will be studied in our future work.

\begin{table*}[t]
\renewcommand{\arraystretch}{1.3}
\center
\begin{tabular}{l}
\toprule
\textbf{Algorithm 1} Flexibly reconfigurable energy-efficient TMA beam steering method\\
\midrule
\textbf{Inputs}: Size of population $N\!P$, Mutation constant $F$, Crossover constant $ C\!R$, Desired Sidelobe level $SLL^{\rm ref}$,\\ \quad \quad \quad Maximum number of iterations $G_{max}$, $+1$st harmonic direction $\theta_{+1}$, \\
\quad \quad \quad $-3$rd harmonic direction $\theta_{-3}$ (dual-beam mode) or $-3$rd harmonic direction $\theta'_{-3}$ (arbitrary angle, single-beam mode).\\
1) Set parameters: Static weighted amplitude $a_n$ (Dolph-Chebyshev distribution).\\
2) \textbf{Initialization}: $\sigma_{ni,0}={\rm rand}(0,1), i\in [1,2,\cdots, N\!P],n=1,2,\cdots,N$\\ \quad \quad (here, $\Sigma_{i,0} = [\sigma_{1i,0},\ldots,\sigma_{Ni,0}]$, $i$ represents the position in the population, and 0 represents the 0th generation, i.e., $G = 0$).\\
3) \textbf{for} $0\leq G\leq G_{max}-1$ \\
\qquad Calculate $\varphi_n,\xi_{n}^{on}$ and $\xi_{n}^{of\!f}$ according to (\ref{Equ18}), where $\sigma_n=\sigma_{ni,G}$.\\
\qquad Calculate the power losses according to (\ref{Equ80}), (\ref{Equ84}), (\ref{Equ88}), (\ref{Equ89}) and (\ref{Equ91}).\\
\qquad Determine $\sigma_{ni,G+1}$ from $\sigma_{ni,G}$ according to (\ref{Equ93}) (DE algorithm; see \cite{DE_algorithm}).\\
\quad \ \textbf{end}\\
4) Select the individuals $\sigma_{n,final}$ with the best fitness from $\sigma_{ni,G_{max}},i\in[1,2,\cdots,N\!P]$ according to (\ref{Equ93}).\\
5) Calculate the new $\varphi_n,\xi^{on}_{n}$ and $\xi^{of\!f}_{n}$ according to (\ref{Equ18}), where $\sigma_n=\sigma_{n,final}$.\\
6) \textbf{if} $\theta_{-1}$ and $\theta_{+1}$ are updated\\
\quad \quad Repeat 5) to calculate the new result.\\
\quad \textbf{end}\\
\textbf{Outputs}: Static weighted phase $\varphi_n$, Turn-on time (normalized) $\xi^{on}_{n}$, Turn-off time (normalized) $\xi^{of\!f}_{n}$.\\
\bottomrule
\end{tabular}
\end{table*}

\section{Numerical Results}
In this section, numerical results are provided based on the proposed algorithm. A linear array composed of 16 omnidirectional antennas with $\lambda /2$ spacing is simulated. The TMA modulation frequency is $F_p = 50\ {\rm MHz}$ and the carrier frequency is $F_c = 1\ {\rm GHz}$. We set the direction of the $+1$st harmonic component to $\theta_{+1} = 80^{\circ}$ and that of the $-3$rd component to $\theta_{-3} = 120^{\circ}$.

\subsection{Power loss with the Chebyshev distribution}

\begin{figure}[!t]
\centering
\setlength{\abovecaptionskip}{0pt}
\setlength{\belowcaptionskip}{-15pt}
\includegraphics[width=3.5in]{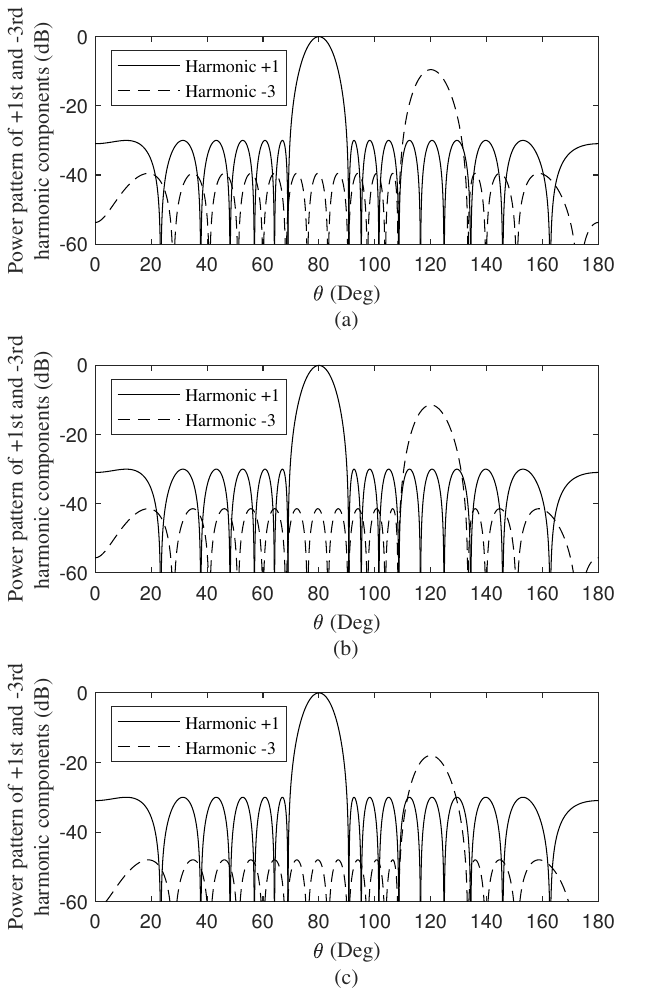}
\caption{Radiation patterns of the TMA (a) with the Chebyshev distribution, (b) in the dual-beam working mode, and (c) in the single-beam working mode.}
\label{Fig6}
\end{figure}

Fig. \ref{Fig6} (a) is plotted as a benchmark; we refer to this case as the Chebyshev distribution, where no energy-efficiency optimization algorithm is used, and the parameters are $\sigma_n = 1,\  n = 1,\ldots,N$. Several indicators are calculated, including the first null beam width (FNBW) $\theta_{+1} = 21.77^{\circ}$ and the sidelobe level $SLL_{+1} = -30\ {\rm dB}$ of the $+1$st harmonic component as well as the FNBW $\theta_{-3} = 25.04^{\circ}$ and the sidelobe level $SLL_{-3} = -39.54\ {\rm dB}$ of the $-3$rd component. The total power loss is $P^{\rm Loss} = 9.94\%$ in the dual-beam mode and $P^{\rm Loss} = 18.94\%$ in the single-beam mode.

\subsection{Power loss in the dual-beam operating mode}
\label{Sec4B}
The simulation results based on the proposed algorithm with power loss optimization for the dual-beam operating mode are shown in Fig. \ref{Fig6} (b). The parameters are configured as follows: $\Sigma = [\sigma_1, \ldots, \sigma_N]$ is the set of parameters to be optimized; the number of dimensions is $D = 16$; the mutation and crossover constants are $F = 0.4$ and $CR = 0.5$, respectively; and the population size is $N\!P = 5D$. The reference sidelobe level is $SLL^{\rm ref} = -30\ {\rm dB}$, and the weighting coefficients are set to $W_{\rm SLL} = 1$, $W_{\rm Loss} = 10$. Fig. \ref{Fig6} (b) shows the radiation pattern after 500 iterations. Several indicators are calculated for comparison, including $\theta_{+1} = 21.77^{\circ}$, $SLL_{+1} = -30\ {\rm dB}$, $\theta_{-3} = 25.21^{\circ}$, and $SLL_{-3} = -41.44\ {\rm dB}$. The total power loss is $P^{\rm Loss} = 3.69\%$. Compared to Fig. \ref{Fig6} (a), the energy efficiency is improved, while the TMA radiation pattern remains the same.

Compared to \cite{JSTSP1}, this paper achieves a reduction in energy loss (from $7.74\%$ to $3.69\%$) at the cost of increased RF front-end structural complexity (due to the additional in-phase component channels and quadrature component channels). In \cite{JSTSP1}, the peak power levels of the two beams are equal, while in this paper, the power levels are different (the difference between the peak values of the +1st and -3rd harmonic components is approximately 11 dB); the latter case is more suitable for scenarios of unequal power, such as high-power radar detection and low-power communication.

\begin{table*}[t]
\renewcommand{\arraystretch}{1.3}
\caption {Summary of important indicators}
\center
\begin{tabular}{c c c c c c c}
\toprule
    & FNBW($\theta_{+1}$) & FNBW($\theta_{-3}$) & $SLL_{+1}$ & $SLL_{-3}$ & $P^{\rm{Loss}}$(Dual-Beam) & $P^{\rm {Loss}}$(Single-Beam)\\
\midrule
Chebyshev Distribution & $21.77^{\circ}$ & $25.04^{\circ}$ & -30\ dB & -39.54\ dB & $9.94\%$ & $18.94\%$\\
Proposed in \cite{JSTSP1} & $21.77^{\circ}$ & / & -30\ dB & / & \boxed{$7.74\%$} & / \\
Proposed Dual-beam mode & $21.77^{\circ}$ & $25.21^{\circ}$ & -30\ dB & -41.44\ dB & \boxed{$3.69\%$} & / \\
Proposed Single-beam mode & $21.77^{\circ}$ & $25.04^{\circ}$ & -30\ dB & -47.99\ dB & / & 7.74\% \\
\bottomrule
\end{tabular}
\label{Tab2}
\end{table*}

\subsection{Power loss in the single-beam operating mode}
The simulation results based on the proposed algorithm with power loss optimization for the single-beam operating mode are shown in Fig.~\ref{Fig6}~(c). The parameters $D$, $F$, $CR$, $N\!P$, $SLL^{\rm ref}$, $W_{\rm SLL}$ and $W_{\rm Loss}$ are the same as in subsection \ref{Sec4B}. Fig. \ref{Fig6} (c) shows the radiation pattern after 500 iterations. Several indicators are calculated for comparison, including $\theta_{+1} = 21.77^{\circ}$, $SLL_{+1} = -30\ {\rm dB}$, $\theta_{-3} = 25.04^{\circ}$, and $SLL_{-3} = -47.99\ {\rm dB}$. The total power loss is $P^{\rm Loss} = 7.74\%$. Compared to Fig. \ref{Fig6} (a), the energy efficiency is improved, while the TMA radiation pattern ($+1$st harmonic component) remains the same. Table \ref{Tab2} summarizes several important metrics for ease of comparison.

\begin{figure}[!t]
\centering
\setlength{\abovecaptionskip}{0pt}
\setlength{\belowcaptionskip}{-15pt}
\includegraphics[width=2.7in]{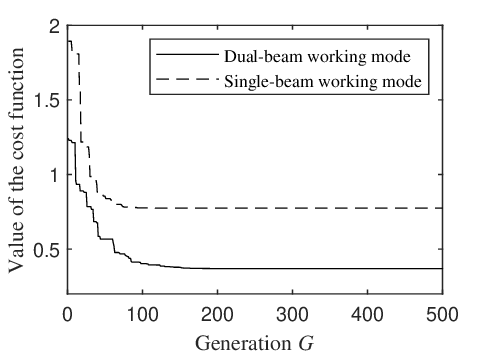}
\caption{Value of the cost function versus generation $G$ for the two working modes.}
\label{Fig7}
\end{figure}
Fig. \ref{Fig7} shows the cost function curves of the DE algorithm versus the number of iterations in the two operating modes. It can be observed that in the single-beam operating mode, the algorithm tends to level out after approximately 100 iterations, and in the dual-beam operating mode, the floor is reached in approximately 160 iterations. Thus, the proposed algorithm has good convergence performance.

\subsection{Comparison with other JCR or TMA systems}
In this section, the energy efficiency is compared both to other TMA-based JCR systems and to other pure TMA energy efficiency optimization design options.

\subsubsection{TMA-based JCR systems}
In the literature \cite{TMA6,TMA7,TMA8,TMA9,TMA10}, the core issue is the realization of the joint communication-radar function, but the energy efficiency of TMA is not considered. Hence its energy efficiency is far from being lower than that of our scheme.

\subsubsection{Pure energy efficient TMA scheme}
The energy efficiency of some state-of-the-art TMAs is compared in Table \ref{Tab3}. It can be observed that the energy efficiency of our solution is comparable to that of the pure TMA design options dedicated to energy efficiency or sideband radiation optimization \cite{EE10,EE13,EE14,EE15,EE22,EE0}. However, these contributions usually employ complex modulation modes or feed structures, and are basically single-beam systems (as shown in Table \ref{TabA1}). Our proposed scheme optimizes energy efficiency while realizing flexible and reconfigurable energy-efficient single/dual-beam TMA, which strikes a compelling compromise in terms of performance, complexity, and energy efficiency.

\begin{table}[t]
\renewcommand{\arraystretch}{1.3}
\caption {Energy efficiency comparison with state-of-the-art TMAs}
\center
\resizebox{0.5\textwidth}{!}{
\begin{tabular}{c|c c c c c c c}
\toprule
    Ref. & \cite{EE10} & \cite{EE13} & \cite{EE14} & \cite{EE15} & \cite{EE22} & \cite{EE0} & Proposed\\
\midrule
$P^{\rm Loss}$ & $75\%$ & $3\%$ & $5.04\%$ & \thead{$4\%$ \\ $9\%$} & \thead{$8.80\%$ \\ $1.28\%$ \\ $0.32\%$} & \thead{$49.88\%$ \\ $67.49\%$ \\ $74.00\%$} & \thead{$3.69\%$ \\ $7.74\%$}\\
\bottomrule
\end{tabular}
}
\label{Tab3}
\end{table}

\section{Summary and Conclusions}
The TMA power loss was investigated in JCR systems, where a single-sideband structure, including an in-phase branch and a $1/4$-cycle delayed quadrature branch, is considered. We first presented some background on TMA beam steering. Then, a closed-form expression was derived for energy efficiency. Based on the result derived, we proposed a flexibly reconfigurable energy-efficient TMA beam steering method, which facilitated the implementation of a single/dual-beam TMA without hardware modification. In the proposed method, we can achieve optimized energy efficiency and arbitrary beam-pointing control (single/dual-beam operating mode). The specific design strategy was given by Algorithm 1. The simulation results were based on a 16-element linear TMA, which confirmed the accuracy of the theoretical derivations. The results showed that in the dual-beam operating mode, the proposed TMA-based JCR system achieved improved energy efficiency at the cost of increased hardware complexity on the RF side compared to \cite{JSTSP1}.

There are still many issues worth studying in future research. For example, consider the resource allocation problems arising from the asymmetry of communications and radar services; optimize the modulation modes and feed structures to further improve energy efficiency of the TMA; design a reconfigurable TMA beam steering method for more beams.

\appendices
\counterwithin{equation}{section}
\counterwithin{figure}{section}
\counterwithin{table}{section}

\section{Proof of equation (\ref{Equ29})}
\label{AppendixA}
We split (\ref{Equ29}) into several parts:
\begin{equation}\label{EquA1}
  \sum_{k=-\infty}^{\infty}|\alpha_{ni,k}|^2 = 2\sum_{k=1}^{\infty}|\alpha_{ni,k}|^2= f_1-f_2-f_3+f_4+f_5,
\end{equation}
\noindent where,
\begin{align}
    f_1 =& 2\sum_{k=1}^{\infty}\frac{1}{\pi^2 k^2} = \frac{1}{3} \tag{\ref{EquA1}a},\\
    f_2 =& 2\sum_{k=1}^{\infty}\frac{\cos(k\pi)}{\pi^2 k^2} = -\frac{1}{6} \tag{\ref{EquA1}b}, \\
    f_3 =& 2\sum_{k=1}^{\infty}\frac{\cos\left(k\cdot 2\pi\left(\xi_{n}^{off}-\xi_{n}^{on}\right)\right)}{\pi^2 k^2} \nonumber
    \\=& \frac{1}{3}-2(\xi_{n}^{off}-\xi_{n}^{on})+2(\xi_{n}^{off}-\xi_{n}^{on})^2 \tag{\ref{EquA1}c}, \\
    f_4 =& 2\sum_{k=1}^{\infty}\frac{\cos\left(k\cdot\left(\pi+2\pi\left(\xi_{n}^{off}-\xi_{n}^{on}\right)\right)\right)}{2\pi^2 k^2} \nonumber
    \\=& -\frac{1}{12} + (\xi_{n}^{off}-\xi_{n}^{on})^2 \tag{\ref{EquA1}d}, \\
    f_5 =& 2\sum_{k=1}^{\infty}\frac{\cos\left(k\cdot\left(\pi-2\pi\left(\xi_{n}^{off}-\xi_{n}^{on}\right)\right)\right)}{2\pi^2 k^2} \nonumber
    \\=& -\frac{1}{12} + (\xi_{n}^{off}-\xi_{n}^{on})^2. \tag{\ref{EquA1}e}
\end{align}
\noindent The derivation of (\ref{EquA1}a)-(\ref{EquA1}e) uses the fact that \cite{ref_table1}
\begin{equation}\label{Equ30}
  \sum_{k=1}^{\infty}\frac{\cos(k\pi)}{k^2} = \frac{\pi^2}{6}-\frac{\pi x}{2} + \frac{x^2}{4},\ x\in[0,2\pi].
\end{equation}

\section{Proof of equation (\ref{Equ35})}
\label{AppendixB}
We analyze the term $|\alpha_{ni,k}|^2\cos\frac{\pi k}{2}$ of (\ref{Equ35}):
\begin{equation}\label{Equ36}
  \begin{aligned}
    |\alpha_{ni,k}|^2\cos\frac{\pi k}{2} =& \frac{\cos\left(\frac{\pi k}{2}\right)}{2\pi ^2 k^2} - \frac{\cos\left(\frac{3\pi k}{2}\right)}{2\pi ^2 k^2}\\
    &-\frac{\cos\left(k\left(\frac{\pi}{2}+2\pi\left(\xi_{n}^{off}-\xi_{n}^{on}\right)\right)\right)}{4\pi^2 k^2} \\
    &-\frac{\cos\left(k\left(\frac{\pi}{2}-2\pi\left(\xi_{n}^{off}-\xi_{n}^{on}\right)\right)\right)}{4\pi^2 k^2} \\
    &+\frac{\cos\left(k\left(\frac{3\pi}{2}+2\pi\left(\xi_{n}^{off}-\xi_{n}^{on}\right)\right)\right)}{4\pi^2 k^2} \\
    &+\frac{\cos\left(k\left(\frac{3\pi}{2}-2\pi\left(\xi_{n}^{off}-\xi_{n}^{on}\right)\right)\right)}{4\pi^2 k^2}
  \end{aligned}
\end{equation}
\noindent Before further derivation, we give the extended result of (\ref{Equ30}):
\begin{equation}\label{Equ37}
  \sum_{k=1}^{\infty}\frac{\cos kx}{k^2} =\left\{
  \begin{aligned}
    & \frac{\pi^2}{6}+\frac{\pi x}{2}+\frac{x^2}{4},\ x\in[-2\pi,0) \\
    & \frac{\pi^2}{6} - \frac{\pi x}{2} + \frac{x^2}{4},\ x\in[0,2\pi] \\
    & \frac{\pi^2}{6} - \frac{3\pi x}{2} + \frac{x^2}{4} + 2\pi^2,\ x\in(2\pi,4\pi]
  \end{aligned}\right.
\end{equation}

By analyzing the six terms in (\ref{Equ36}) separately, we obtain
\begin{equation}\label{Equ38}
  f'_1 = 2\sum_{k = 1}^{\infty}\frac{\cos\left(\frac{\pi k}{2}\right)}{2\pi^2 k^2} = -\frac{1}{48},
\end{equation}
\begin{equation}\label{Equ39}
  f'_2 = 2\sum_{k=1}^{\infty}\frac{\cos\left(\frac{3\pi k}{2}\right)}{2\pi^2 k^2} = -\frac{1}{48},
\end{equation}
\begin{equation}\label{Equ40}
\begin{aligned}
  f'_3 =& 2\sum_{k=1}^{\infty}\frac{\cos\left(k\left(\frac{\pi}{2}+2\pi\left(\xi_{n}^{off}-\xi_{n}^{on}\right)\right)\right)}{4\pi^2 k^2}
  = -\frac{1}{96}-\frac{\tau_n}{4}+\frac{\tau_n^2}{2},
\end{aligned}
\end{equation}
\begin{equation}\label{Equ41}
  \begin{aligned}
  f'_6 =& 2\sum_{k=1}^{\infty}\frac{\cos\left(k\left(\frac{3\pi}{2}-2\pi\left(\xi_{n}^{off}-\xi_{n}^{on}\right)\right)\right)}{4\pi^2 k^2}
  = -\frac{1}{96}-\frac{\tau_n}{4} + \frac{\tau_n^2}{2}.
  \end{aligned}
\end{equation}

For the fourth and fifth terms of (\ref{Equ36}), we discuss them in different cases. If $\tau_n \in [0,\frac{1}{4}]$, then
\begin{equation}\label{Equ42}
  \begin{aligned}
    f'_4 =& 2\sum_{k=1}^{\infty}\frac{\cos\left(k\left(\frac{\pi}{2}-2\pi\left(\xi_{n}^{off}-\xi_{n}^{on}\right)\right)\right)}{4\pi^2 k^2}
    = -\frac{1}{96} + \frac{\tau_n}{4} + \frac{\tau_n^2}{2},
  \end{aligned}
\end{equation}
\begin{equation}\label{Equ43}
  \begin{aligned}
    f'_5 =& 2\sum_{k=1}^{\infty}\frac{\cos\left(k\left(\frac{3\pi}{2}+2\pi\left(\xi_{n}^{off}-\xi_{n}^{on}\right)\right)\right)}{4\pi^2 k^2}
    = -\frac{1}{96} + \frac{\tau_n}{4} + \frac{\tau_n^2}{2}.
  \end{aligned}
\end{equation}
\noindent If $\tau \in (\frac{1}{4},\frac{1}{2}]$, then
\begin{equation}\label{Equ44}
  \begin{aligned}
    f'_4 =& 2\sum_{k=1}^{\infty}\frac{\cos\left(k\left(\frac{\pi}{2}-2\pi\left(\xi_{n}^{off}-\xi_{n}^{on}\right)\right)\right)}{4\pi^2 k^2}
    = \frac{23}{96} - \frac{3\tau_n}{4} + \frac{\tau_n^2}{2},
  \end{aligned}
\end{equation}
\begin{equation}\label{Equ45}
  \begin{aligned}
    f'_5 =& 2\sum_{k=1}^{\infty}\frac{\cos\left(k\left(\frac{3\pi}{2}+2\pi\left(\xi_{n}^{off}-\xi_{n}^{on}\right)\right)\right)}{4\pi^2 k^2}
    = \frac{23}{96} - \frac{3\tau_n}{4} + \frac{\tau_n^2}{2}.
  \end{aligned}
\end{equation}
\noindent In either case, we have $-f'_4+f'_5 = 0$. Then,
\begin{equation}\label{Equ46}
  \sum_{k=-\infty}^{\infty}\alpha_{ni,k}\alpha_{nq,k}^{\ast} = f'_1 - f'_2 - f'_3 -f'_4 + f'_5 +f'_6 =0.
\end{equation}

\section{Proof of equation (\ref{Equ64})}
\label{AppendixC}
\begin{figure*}[!t]
\centering
\includegraphics[width=6.5in]{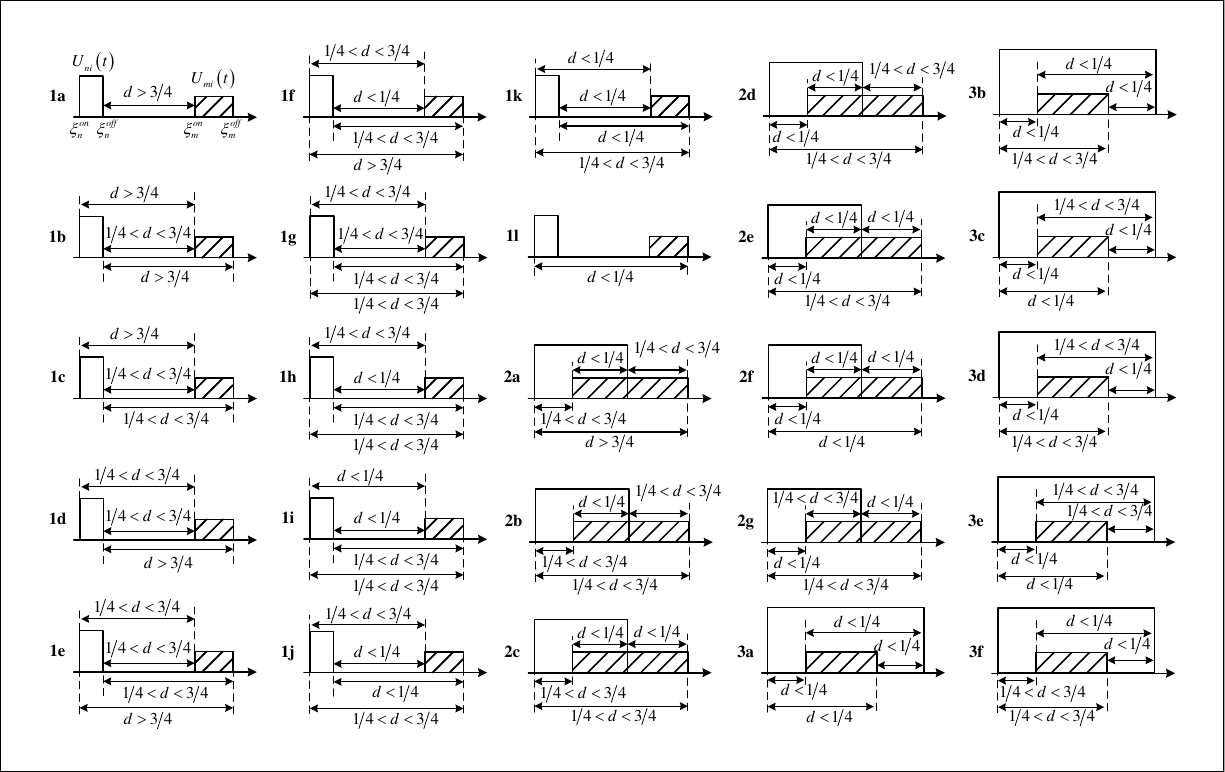}
\caption{Three types of cases of the positional relationship between pulses $m$ and $n$.}
\label{Fig4}
\end{figure*}

According to reference \cite{ref_table1}, we have
\begin{equation}\label{Equ60}
  \sum_{k=0}^{\infty}\frac{\sin(2k+1)x}{(2k+1)^2} = \left\{
  \begin{aligned}
    &\frac{\pi x}{4},&x\in[-\frac{\pi}{2},\frac{\pi}{2}) \\
    &\frac{\pi (\pi-x)}{4},&x\in[\frac{\pi}{2},\frac{3\pi}{2}]
  \end{aligned}\right.
\end{equation}
\noindent Because $\xi_a - \xi_b \in [-1,1]$, where $\xi_a$ represents $\xi^{on}_m$ or $\xi^{off}_m$ and $\xi_b$ represents $\xi^{on}_n$ or $\xi^{off}_n$, we extend (\ref{Equ60}) to $x\in[-2\pi,2\pi]$. After some simple derivation,
\begin{equation}\label{Equ61}
  \sum_{k=0}^{\infty}(-1)^k\frac{\sin(2k+1)x}{(2k+1)^2}=\left\{ \\
  \begin{aligned}
    &\frac{\pi(x+2\pi)}{4},&x\in[-2\pi,-\frac{3\pi}{2}) \\
    &-\frac{\pi(x+\pi)}{4},&x\in[-\frac{3\pi}{2},-\frac{\pi}{2}) \\
    &\frac{\pi x}{4},&x\in[-\frac{\pi}{2},\frac{\pi}{2}) \\
    &\frac{\pi(\pi -x)}{4},&x\in [\frac{\pi}{2},\frac{3\pi}{2}) \\
    &\frac{\pi(x-2\pi)}{4},&x\in[\frac{3\pi}{2},2\pi]
  \end{aligned}\right.
\end{equation}

In accordance with the position relationship of $U_{mi}(t)$ and $U_{ni}(t)$ within a period, the different possible situations are discussed separately. For succinctness of expression, the variables $U_{mi}(t), U_{ni}(t), \xi_{m}^{on},\xi_{n}^{on},\xi_{m}^{off}$ and $\xi_{n}^{off}$ are shown only in case 1a of Fig. \ref{Fig4}, where $d$ represents the length (normalized with respect to the period $T_p$). In cases 1a to 1l, pulse $n$ is on the left side of pulse $m$, and there is no overlap between the two pulses. In cases 2a to 2g, pulse $n$ is on the left side of pulse $m$, and the two pulses overlap. In cases 3a to 3f, pulse $n$ entirely contains pulse $m$. The cases in which pulse $m$ is on the left side of pulse $n$ and in which $m$ contains $n$ are not listed because the results for these cases can be readily deduced from those for the above cases (1a to 3f). According to (\ref{Equ59}), (\ref{Equ61}) and Fig. \ref{Fig4}, we can calculate the value of $\sum_{k=-\infty}^{\infty}\alpha_{ni,k}\alpha_{mq,k}^{\ast}$ based on the different cases.

We take case 1f as an example for detailed description. In this case,
\begin{equation}\label{Equ62}
  \begin{aligned}
    \sum_{k=-\infty}^{\infty} \alpha_{ni,k}\alpha_{mq,k}^{\ast} =& \frac{2}{\pi^2}\Big(\frac{\pi}{4}\left(\pi-2\pi\left(\xi_{m}^{on}-\xi_{n}^{on}\right)\right)
    \\&+\frac{\pi}{4}\left(\pi-2\pi\left(\xi_{m}^{off}-\xi_{n}^{off}\right)\right) \\
    & -\frac{\pi}{4}\left(2\pi\left(\xi_{n}^{off}-\xi_{n}^{on}\right)-2\pi\right)
    \\& -\frac{\pi}{4}2\pi\left(\xi_{m}^{on}-\xi_{n}^{off}\right)\Big) \\
    =&2\big(\xi_{n}^{off}-\xi_{m}^{on}-\xi_{m}^{off} +\xi_{n}^{on}+1\big).
  \end{aligned}
\end{equation}
\noindent In addition, we find that (as shown in Fig. \ref{Fig5})
\begin{equation}\label{Equ63}
  \begin{aligned}
    \sum_{k=-\infty}^{\infty} \alpha_{ni,k}\alpha_{mq,k}^{\ast} =&2\Big(\big(\xi_{n}^{off}-(\xi_{m}^{on}-\frac{1}{4})\big)
    \\&-\big((\xi_{m}^{off}-\frac{1}{4})-(\xi_{m}^{on}+\frac{1}{4})\Big)
    \\=& \tau'_+ - \tau'_-,
  \end{aligned}
\end{equation}

\noindent Similarly, all other cases (case 1a to case 3f and the cases not shown in Fig. \ref{Fig4}) can be analyzed as above, and the same result can be obtained.

\normalem
\bibliographystyle{./IEEEtran}
\balance
\bibliography{./IEEEabrv,./IEEEexample}

\end{document}